# Comments on "Wall-plug (AC) power consumption of a very high energy e+/e- storage ring collider" by Marc Ross [arXiv:1308.0735]

**A. Blondel, M. Koratzinos,** U. Geneva, Switzerland;

**A. Butterworth, P. Janot, F. Zimmermann,** CERN, Switzerland;

**R. Aleksan,** CEA, France;

**P. Azzi,** INFN Padova Italy;

**J. Ellis,** King's College, London, UK & CERN, Switzerland;

**M. Klute** and **M. Zanetti,** MIT, USA

### Abstract

*The paper arXiv:1308.0735 questions some of the technical assumptions made by the TLEP Steering Group when estimating in arXiv:1305.6498 the power requirement for the very high energy $e^+e^-$ storage ring collider TLEP. We show that our assumptions are based solidly on CERN experience with LEP and the LHC, as well accelerators elsewhere, and confirm our earlier baseline estimate of the TLEP power consumption.*

### Introduction

In [1], we estimated that the wall-plug power consumption for TLEP, an $e^+e^-$ storage ring collider with $E_{cm}$ = 350 GeV and 80 km circumference, would be about 280 MW, whereas in [2], M. Ross estimated about 416 MW. As he explained in [3], the main differences between the two power estimates arise from different assumptions about klystron operation, cryo-plant efficiency and heat removal. Our technical assumptions concerning these and other issues raised in [2,3] are based solidly on CERN experience with LEP and the LHC, as well accelerators elsewhere. In this note, we summarize the bases for our technical assumptions: we see no reason to modify significantly the wall-plug power consumption for TLEP estimated in [1]. In the following, we address point-by-point the issues raised in [3].

### 1) Klystron operation

Proposing a reduction of the klystron efficiency assumed for TLEP in [1] from 65% to 55%, it is stated in [3] that "*'Saturated' klystron operation is very unusual in storage rings. At LEP this was done only for the last year of operation in an attempt to capture as much luminosity at the highest achievable energy and is not a reasonable approach to take for a new machine. During that last LEP-year, any perturbation generated a 'beam-trip'*".

These statements are inaccurate [4]. The majority of the LEP klystrons were operated in saturation, from the very beginning of LEP1 operation. A slow amplitude control loop acted on the klystron modulation anode. There was no fast feedback. It was done this way for efficiency reasons. The sensitivity to trips in the last year of LEP operation was NOT due to running at saturation, it was due to the fact that the beam energy was pushed to a level where we had very little voltage margin and many cavities were running at their absolute maximum gradient.

During LEP2, a fast vector sum feedback was tried with the aim of raising the threshold of beam instabilities at low energy, but this was found to be too sensitive to perturbations and resulted in too many RF trips and lost fills. Ultimately only a few RF units were ever operated using this feedback and the vast majority of the units continued to operate using the slow scalar voltage control.

We see, therefore, no reason to reduce the klystron efficiency from 65% to 55% as suggested in [3].

The estimate made for TLEP in this context is actually quite conservative. The efficiency of the RF power source is the single most important source of power losses, on which dedicated component R&D is explicitly planned at CERN. The aim is to improve the efficiency significantly, from the assumed 54% up to 70-80%. It is also planned to study possible heat recovery mechanisms. Neither of these potential savings have been included in the estimates so far.

**2) Cryo-plant efficiency**

Proposing a reduction of the cryo-plant efficiency assumed for TLEP in [1], it is stated in [3] that the *"cryo–plant power required at JLab for 1.9 degrees K is 1100 W per Watt dissipated at low temperature [which] is 20% worse than the assumed value in (1)"*.

The efficiency we assumed for the cryogenic system is what the LHC cryo-system achieves, namely 900W/W at 1.9 K, which we consider to be a reasonable estimate for TLEP [1]. This assumption turns out to be conservative: we note in passing that the ILC TDR design [5] is significantly more aggressive in this respect, as it assumes 700W/W, which is 29% better than the LHC and our assumption for TLEP, and 57% better than what JLab achieves.

We see, therefore, no reason to either increase or decrease our estimate of the cryo-plant efficiency.

**3) Ventilation System**

Proposing an increase of the power of the ventilation system, it is stated in [3] that *"Following the Swiss Alps [road] tunnel fires and the LHC cryogenic incident CERN decided to adopt an actively controlled 'transversal' tunnel ventilation system [6]. This is much safer as it allows*

*segmented control of tunnel air flow but it is more expensive and requires more extensive ventilation equipment".*

Contrary to this statement, CERN has not decided to have transversal tunnel ventilation anywhere, with the possible exception of CLIC [7]. The reason in the CLIC case would not be safety constraints (which are very different from those in a road tunnel) but because of the expected high heat load to air. The fact that this improves the situation from the fire safety point of view is not crucial for the choice. Even for CLIC, this decision is not yet final, however, since the design for CLIC requires additional work to validate some parameters that are only preliminary at this moment.

In the case of TLEP, most of the SR heat load should be on the water cooling for the photon stops and not on air, so the CLIC argument would anyway not apply.

We see, therefore, no reason to change our estimate of the power requirement of the ventilation system.

**4) Water cooling**

It is stated in [3] that *"The electrical power required to remove heat through primary and secondary water cooling system loops in typical accelerator installations is between 5 and 10% .... For magnet systems we should assume 10% is required."*

It is clear that the water cooling system for TLEP will need to be designed carefully. To compute the total power, we would need a concrete design for the photon stops and associated cooling system, which is not available at the moment. There is also interplay between the system cost and its power consumption. Our estimate of 5 MW was estimated from first principles, using figures for power consumption for existing CERN cooling equipment.

In our estimate [1], the electric power for cooling and ventilation represents about 10% of the total electric power, a figure which we will strive to achieve with a careful design. For comparison, in the last year of LEP operation (2000), the LEP cooling and ventilation together amounted to 13% of the total LEP electric power consumption [8]. LEP was designed and built in the early 1980s. If we assumed the same percentage for TLEP we would need to increase the estimate of [1] by about 10 MW.

**5) Electrical distribution network losses**

It is stated in [3] that *"Electrical network losses, which may be substantial for a distributed complex, will be about 5% "*.

Indeed, in our estimate we did not include any electrical network losses, as they are dependent on the exact implementation of the electrical distribution system. Efforts will be made to ensure that the main electrical power consumers (the RF systems) will be located close to local utility high-voltage interconnect points. We consider that 5% is on the conservative side and we strive to achieve a figure closer to 3% for the electrical distribution network [9].

**6) RF system**

It is stated in [3] that *"Above about 100 mA (9% of nominal 90 GeV $E_{cm}$) multi-cell RF cavities are not used because of trapped higher-order-modes [10]. Both PEP-II and KEK-B use heavily damped single-cell RF cavities with a packing-factor 5 to 10 times worse than multi-cell cavities. The RF effective length then becomes 3 to 6 km and the cost of the SRF system would scale accordingly."*

We do not see any such statement in Ref. [10].

We note that three of the few high-energy lepton colliders operated so far with SC RF used multi-cell cavities (TRISTAN 14.5 mA, LEP 8.5 mA, and HERAe ~100 mA]. As far as we are aware the beam currents in all of these machines were not limited by the multi-cell nature of the RF cavities, but mainly by the SR power.

The planned eRHIC project with 600 mA beam current in the baseline design [11] (not the 300mA quoted in [10, 2]) will use multi-cell cavities very similar to the cavities considered for TLEP. The cavity design (main power coupler, HOMs) has been optimized so as to cope with these beam currents. The multi-cell cavities for the BNL test ERL, under construction, are designed for a beam current of 1 A [12].

We also note that in 2005 a 7-cell 1.3 GHz SC RF cavity was designed for the DAFNE collider to operate with beam currents above 1 A [13].

We see, therefore, no reason to change our estimate of the length and cost of the RF system.

**7) Magnet Power consumption**

It is stated in [3] that *"For $E_{CM}$ = 350 GeV, there is no valid collider ring lattice design. We can assume the number of cells in the lattice to be substantially increased, compared to LEP or LHeC, in order to achieve the needed momentum acceptance of 2.5%. For modern synchrotron light ring optics this can be about a factor of two to five. This factor, applied to two instead of one ring, gives 12 times (or 30 times) as many magnets as listed in the LHeC design giving a ring power consumption of 42 MW, 3 times larger than the listed value of 14 MW."*

This statement is inaccurate. First, the momentum acceptance limitation is not in the ring but in the low-beta insertion region: the arcs have very large momentum acceptance. The power calculation considered a lattice similar to the LHeC study and magnets based on aluminum conductor. Secondly, a lattice exists for $E_{CM}$ = 350 GeV, generated with a similar cell length as that of the LHeC. It yields an emittance slightly smaller than desired, because the dipole field is so small. This result indicates that the number of quadrupoles could be somewhat *reduced*, i.e., the cells lengthened, which would actually *decrease* the power consumption from what we have assumed. For operation at lower energies, the optical cells need to be lengthened by factors 2 or 6, respectively, with a corresponding decrease in the number of active quadrupoles and in the corresponding electrical power. The fact that TLEP requires separate channels for $e^+$ and $e^-$, in contrast to the LHeC, will increase the power consumption. However, the use of copper instead of aluminum as magnet conductor would decrease the estimated consumption by a comparable factor. More precise numbers will be provided when complete definitions of the lattice and the magnets are available.

**Final remarks**

The baseline TLEP design aims at a total power consumption below 300 MW, and the arguments given in [2,3] do not lead us to abandon this objective. We note, however, that for a given optics, the luminosity of a circular collider scales linearly with the SR power dissipated in the arcs and is, therefore, almost directly proportional to the total electric power. Thus a trade-off between power consumption and luminosity is in principle possible, if necessary. We consider that the estimate of the TLEP power consumption given in [1] is a good starting point for the TLEP design study planned for the coming two years. It will aim to reduce the total power needed as much as possible, e.g., by increasing the RF power efficiency as well as possibly by recycling the SR heat dissipated in the arcs.

**Acknowledgements**

We thank M. Ross for his interest and careful consideration of the TLEP project, and we look forward to further exchanges with him and others.